\documentclass{aastex}
\usepackage{emulateapj5}

\slugcomment{Accepted to the Astrophysical Journal} 

\shorttitle{A Photoionized Plasma Above the Accretion Disk in EXO 0748-676}
\shortauthors{Jimenez-Garate et al.}

\def\exo{EXO~0748-676 }
\def\exop{EXO~0748-676}
\def\chandra{{\it Chandra }}
\def\xmm{{\it XMM-Newton }}
\def\asca{{\it ASCA }}
\def\rosat{{\it ROSAT }}

\def\degr{$^{\circ}$}

\def\cm{{\rm\thinspace cm}}
\def\counts{{\rm\thinspace counts}}

\def\erg{{\rm\thinspace erg}}
\def\eV{{\rm\thinspace eV}}
\def\g{{\rm\thinspace g}}

\def\keV{{\rm\thinspace keV}}
\def\km{{\rm\thinspace km}}
\def\kpc{{\rm\thinspace kpc}}

\def\Msun{\hbox{$\rm\thinspace M_{\odot}$}}

\def\ph{{\rm\thinspace ph}}
\def\s{{\rm\thinspace s}}

\def\cps{\hbox{$\counts~\s^{-1}\,$}}

\def\pcmcu{\hbox{$\cm^{-3}\,$}}

\def\ergcmps{\hbox{$\erg\cm\ps\,$}}

\def\ergpcmsqpspA{\hbox{$\erg\cm^{-2}\s^{-1}$\AA$^{-1}\,$}}
\def\ergps{\hbox{$\erg\s^{-1}\,$}}
\def\ergpspa{\hbox{$\erg\s^{-1}$\AA$^{-1}\,$}}

\def\gps{\hbox{$\g\s^{-1}\,$}}
\def\kmps{\hbox{$\km\s^{-1}\,$}}

\def\pcm{\hbox{$\cm^{-3}\,$}}
\def\pcmsq{\hbox{$\cm^{-2}\,$}}

\def\phpcmsqps{\hbox{$\ph\cm^{-2}\s^{-1}\,$}}

\def\ps{\hbox{$\s^{-1}\,$}}

\def\powerlawfluxat1kev{\hbox{$\ph\cm^{-2}\s^{-1}\keV^{-1}$}}

\def\nH{$N_{\rm H}\,$}

\begin{document}


\title{Discrete X-ray Signatures of a Photoionized Plasma
Above the Accretion Disk of the Neutron Star EXO 0748-676}


\author{M. A. Jimenez-Garate}
\author{N. S. Schulz}
\author{H. L. Marshall}
\affil{MIT Center for Space Research, 70 Vassar St, NE80 6th floor, Cambridge, MA 02139}
\email{mario@alum.mit.edu,nss@space.mit.edu,hermanm@space.mit.edu}



\begin{abstract}
During the disk-mediated accretion phase, 
the high-resolution X-ray spectrum of 
the low-mass X-ray binary system EXO 0748-676 reveals
a photoionized plasma which is orbiting the neutron star.
Our observations with the \chandra High Energy
Transmission Grating Spectrometer (HETGS) 
constrain the structure of the upper layers of the accretion disk,
by means of the recombination emission lines from the H-like and He-like
ions of O, Ne, and Mg, which have a mean velocity broadening
$\sigma_v \sim 750 \pm 120$~\kmps. 
The \ion{Mg}{11} emission region has density $n_e \gtrsim 10^{12}$\pcm and 
is located within $7 \times 10^{9} < r < 6 \times 10^{10}$\cm \ of the neutron star,
while the temperature of the \ion{Ne}{10} region is $kT \lesssim 20$\eV.
These lines favor a vertically stratified distribution of ions in the disk.
The spectra show that the line region is spatially extended and unabsorbed, while 
the continuum region is compact and heavily absorbed.
The absorber has variable column density and is
composed of both neutral and ionized gas, which can explain the stochastic
and periodic X-ray intensity dips, the X-ray continuum evolution, and
the \ion{O}{7} and \ion{Mg}{11} K-shell 
absorption edges.  The absorber is located 
8\degr--15\degr \ above the disk midplane,
inclusive of two bulges near the disk edge.
This outer disk gas may participate in
the outflow of ionized plasma which was previously identified in \xmm grating 
spectra obtained during type I bursts.
The thickened photoionized region above the disk can be produced
by heating from the neutron star X-rays and by the impact of the accretion stream.
\end{abstract}




\section{Introduction}
\label{sec:intro}

\exo provides a unique view of a photoionized plasma in the region near
an accretion disk.
\exo is a Low-Mass X-ray Binary (LMXB) which exhibits 
eclipses, type I X-ray bursts, and intensity dips 
\cite[]{discovery, exo_ecl}. The
X-ray eclipses have a period $P_{\rm orb} = 3.82$~hr
and last for 8.3~min. 
Assuming Roche lobe overflow and 
a primary accretor with $M_{NS} \sim 1.4$\Msun, Parmar et al. obtained 
a companion mass of $0.08 < M_{C} < 0.45$\Msun.
They found that the largest $M_{C}$ corresponds to a main-sequence companion, 
with a system inclination of 75\degr$ < i < 83$\degr.
As we will see below, this inclination is favorable
for X-ray spectroscopic studies of the accretion flow.
The type I bursts are due to a thermonuclear
flash on the surface of an accreting neutron star,
which requires a magnetic field $B \lesssim 10^{10}$~G
\cite[and references therein]{lewin}.
The X-ray burst luminosity is near
the Eddington limit, for a
distance to \exo of $D \sim 10$\kpc, so we will assume this value of $D$.
The low $B$ implies that the accretion disk can extend all the way in, 
down to the neutron star surface.
The persistent X-ray luminosity is $L \sim 10^{37}$\ergps,
and most of this power is released in the innermost
radii of the accretion disk and in the boundary layer
of the neutron star.

Intensity dips are often present in LMXB 
observed at high-inclination \cite[]{incl}. 
Frank et al. attributed the X-ray intensity dips 
to absorption from photoionized clouds in a two-phase medium
above the outer accretion disk. 
In \exop, dips are 
associated with X-ray spectral evolution.
The dips are observed to precede eclipses at 
orbital phase $\phi \sim 0.9$, or to occur
during mid-orbit at $\phi \sim 0.6$. The depth and
duty cycle of the dips varies with each orbit. 
During the dips, the spectrum hardens, 
as would be expected if they are caused by increased
photoelectric absorption 
\cite[]{exo_ecl}. 
A soft X-ray excess was observed at 
$8 < \lambda < 30$~\AA \ with \asca and \rosat \cite[]{asca_excess,rosat}. 
This excess was presumed to be associated with the 
existence of two emission components, or with the 
``partial covering'' of the X-ray source by
a rapidly variable low-metallicity absorber
\cite[]{exo_ecl,partcover}. 

High-resolution 10--23~\AA \ spectra of \exo 
showed broad X-ray line emission and drastic spectral variability.
\cite{rgs} found recombination lines from \ion{N}{7}, \ion{O}{7}, \ion{O}{8}, \ion{Ne}{9}, and \ion{Ne}{10},
using the \xmm Reflection Grating Spectrometer (RGS).
The K edges of \ion{O}{7} and \ion{O}{8} were visible 
during the rapid variation periods, but they faded as
the soft continuum dimmed.
The velocity broadening of the recombination lines was
$\sim 10^3$\kmps, and it was found to be correlated with
the ionization parameter. 
C01 concluded that both emission and absorption features
originate from an extended, oblate structure
above the accretion disk. 

A 335~ks-long exposure of \exo obtained with the \xmm RGS also 
revealed that the cumulative spectrum of 28 type I bursts 
contains absorption lines which may be identified
with \ion{Fe}{25} and \ion{Fe}{26} 
at a redshift of $z=0.35$ \cite[]{red}. 
These spectral features could represent the
first measurement of a gravitational redshift
from the surface of a neutron star, constraining $M_{NS}/R_{NS}$ and
the nuclear equation of state at high density.
The spectra of the 5 type I bursts detected with the \chandra High
Energy Transmission Spectrometer (HETGS)
will be presented elsewhere (Marshall et al., in preparation).

In this work, we use the \chandra HETGS to perform a spectroscopic
analysis of \exo during epochs exclusive of bursts
in order to study the circumsource medium of the neutron star, 
and to constrain the elemental abundances of the accreting matter.
The broadband high-resolution spectra of the HETGS, 
from $1.5 < \lambda < 25$~\AA, allows us to identify the
nature of the absorber which produces
the intensity dips. We measure the density, location, and kinematics of
the recombination line region, which constrain models of
accretion disk structure. 

\section{\chandra HETGS Data Reduction}
\label{sec:obs}

The observation was performed on 2001 April 14, with
a 47.7~ks exposure (starting at Modified Julian Day 52013.052).
Our data was obtained with the 
HETGS \cite[]{hetg}, which provides two independent spectra:
the high-energy grating (HEG) spectrum has a resolution of
$\Delta \lambda = 0.012$~\AA \ (FWHM),
and the medium-energy grating (MEG) spectrum has a
$\Delta \lambda = 0.023$~\AA.
We restrict our attention to the 
$1.5 < \lambda < 14$~\AA \
portion of the HEG spectrum and the 
$2.0 < \lambda < 25$~\AA \ portion of the MEG spectrum. 
The photons were detected with the Advanced CCD Imaging
Spectrometer (ACIS-S) detector.
The data were processed to level 2 with the CIAO tools version 2.2.1,
\footnote{CIAO is the {\it Chandra} Interactive Analysis of Observations, 
a software system developed by the {\it Chandra} X-Ray Center.}
using the default spatial and pulse-height extraction
regions for the HEG and MEG spectra.
We used the {\it lightcurve} tool to histogram
the count rate as a function of event time. 
We used the {\it dmcopy} tool
to select time cuts from the event list. 
We produced event lists for two different states which were selected by count rate.
We generated an effective area for each event list selection,
as well as the exposure maps needed to obtain fluxed spectra
using {\it mkgarf}. The rebinning
and fitting of spectra were performed using the ISIS program \cite[]{isis}.
Spectra from the $+1$ and $-1$ grating orders were combined to obtain
the total HEG and MEG spectra.
These spectra are fit simultaneously with phenomenological models.
The HEG and MEG spectra are summed for display purposes only, since their
line response functions differ considerably.

We performed continuum fits on the HETGS spectra and
searched for discrete features.
Our criteria for the identification 
of a discrete spectral feature are that
1) the number of counts in the line 
represents a $>3\sigma$ deviation from the continuum level,
2) the equivalent width (EW) of the feature is consistent in both 
HEG and MEG spectra when both gratings have sufficient counts,
or else that the feature is observed in the $+1$ and $-1$
orders of the grating with the largest effective
area, 3) the feature does
not fall within a chip gap, and 4) the 
identified feature is spectroscopically tenable.
We determined limits on some of the expected features.
The unidentified residuals may, in some
cases, be due to instrumental effects.
Effective area systematic errors are not included in
our plots. Systematic errors are occasionally visible in the 5.4 to 6.2~\AA \
region due to uncertainties at the Ir~M-shell edges.

\section{Light curves}
\label{sec:lightcurve}

The X-ray light curve of \exo exhibits eclipses, 
intensity dips, and type I bursts.
Previous observations of \exo 
with low spectral resolution instruments \cite[]{exo_ecl,asca_excess}
yielded light curves which
resemble the HETGS broadband light curve in
Figure \ref{fig:totlc}. 
We split the HETGS light curve into three wavebands.
The HETGS wavebands have sharp, resolution-limited 
boundaries ($\Delta \lambda = 0.020$~\AA).

The appearance of the light curve varies
from one waveband to another (as shown in Figure \ref{fig:wavelc}). 
Both the behavior of the light curves
and the spectroscopic data presented below
indicate that this variation is
mainly due to intensity dips.
In the 1.5 to 3~\AA \ band, the X-ray flux is nearly constant
outside the eclipse and burst epochs, and only traces of dips are
discernible. 
Most of the variability in the hard band of the persistent 
light curve is statistical. Significant changes in the accretion rate 
are not discernible in the light curve.
We observe four eclipses, each lasting $500 \pm 10$~s. We find
five type I X-ray bursts which are each $t \lesssim 10^2$~s in duration.
At least some of the variation of
burst brightness appears to be due to a variable absorbing column 
of gas in the line-of-sight, 
since the dimmest burst (the fifth in time order) occurs
during a pre-eclipse dip phase.
The second-dimmest burst (also the second in time) occurs just before a
pre-eclipse dip.
In the 3 to 6~\AA \ band, there are 
two types of dips, stochastic and periodic. The periodic dips occur at mid-orbit
($\phi \sim 0.6$) and before eclipse ingress 
($\phi \sim 0.9$). 
The light curve evolves with each successive binary orbit.
The evolution of the dips 
from one binary orbit to the next
indicates that the accretion flow geometry 
is changing at the outer edge of the accretion disk.
Above 6~\AA, the light curve appears to have
``soft flare'' events
\cite[C01]{epic}, one of which is
in our data 
near the 8~ks mark. 
Alternatively, these ``soft flares'' can be explained by the variability of the absorber column,
which is found in the high-resolution spectra (\S \ref{sub:dips}). 
Above 6~\AA, the eclipse ingresses appear to fade away because of 
deep pre-eclipse dips. 
In contrast, the eclipse egresses do appear in the
$\lambda > 6$~\AA \ light curve because dips are shallow
at those times. 

The intensity dip timescale is shorter than the size of the bins in
the HETGS light curve ($<50$~s). There are rapid intensity fluctuations 
which drive the count rate from maximum to minimum, or an order 
of magnitude change in the soft ($\lambda > 6$~\AA) band, as shown 
in Figure \ref{fig:wavelc}.
These fluctuations are common in the data and are statistically significant.

We obtain spectra for selected time intervals,
with cuts based on the light curve, to isolate events during 
the deep dip state, the persistent (or shallow dip) state, 
the eclipses, and the burst events. 
To make the cuts, we use the light curve for $\lambda > 6$~\AA \
in Figure \ref{fig:wavelc}. First, we exclude
the times during bursts and eclipses. 
Then, we set a fiducial cut at 0.3~\cps, 
below which
we obtain the {\it dip state} with an exposure of 29.5~ks, and above which
we obtain the {\it persistent state}, with 15.7~ks. 

\section{High-Resolution X-ray Spectroscopy of Dip and Persistent States}
\label{sub:dips}

\begin{figure}
\epsscale{0.85}
\plotone{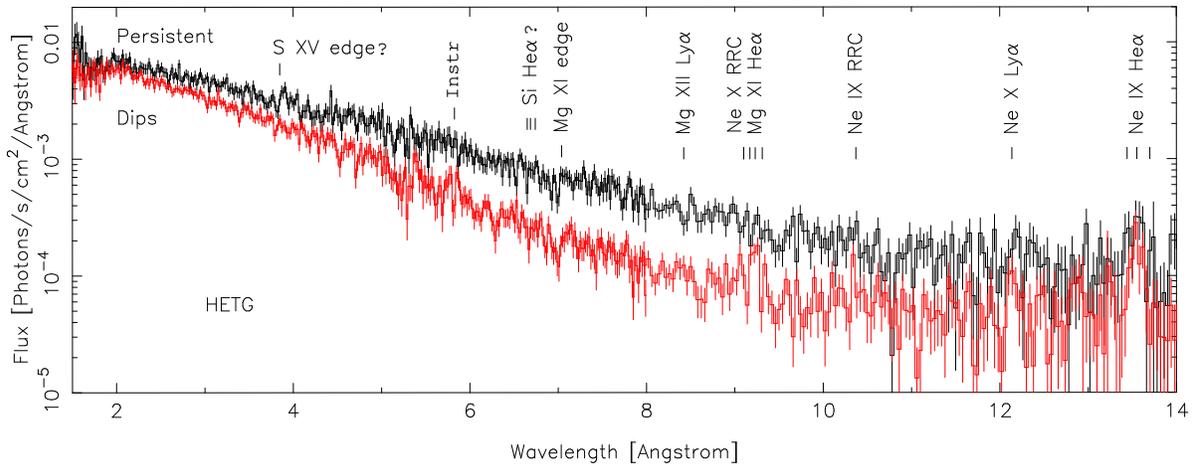}
\caption{Summed MEG and HEG spectra; (top curve) 15.7~ks during the persistent state, 
(bottom curve) 29.5~ks during the dip state. During the dip
phases, the hard continuum component is
absorbed while the soft recombination lines remain unabsorbed.
\label{fig:dip_pers}}
\end{figure}

The HETGS X-ray spectra reveal emission and absorption 
features which evolve with the soft X-ray intensity. 
The EW of the lines and the edge depths 
are larger during the dip state than during the persistent state.
However, the line fluxes remain constant.
The intensity dip spectra are best fit by an absorber
composed of both neutral and ionized gas.
Figure \ref{fig:dip_pers} shows the
combined HEG and MEG spectra.
The spectra require three emission components: 
(1) a bright, hard, power-law continuum;
(2) a dim, soft continuum; and (3) a recombination emission component with
resolved velocity broadening.
Components 2 and 3 have a smaller absorbing column than component 1. 

The observed discrete emission is produced by electron-ion recombination
in a photoionized plasma. The lines detected with the
largest statistical significance, as well as upper limits for selected lines,
are listed in Table \ref{tab:lines}.
The recombination features consist of lines from
H-like and He-like ions plus radiative recombination
continua (RRC). 
The most prominent lines in the HETGS spectra
are the \ion{Mg}{11} and the \ion{Ne}{9} 
intercombination ($i$) lines, and \ion{O}{8} Ly$\alpha$.
\ion{Mg}{12} Ly$\alpha$,
\ion{Ne}{10} Ly$\alpha$, \ion{Ne}{10} RRC, 
\ion{Ne}{9} RRC, and \ion{O}{7} $i+r$ 
are weaker but still detected with a significance $> 3 \sigma$.
Possible detections include
\ion{Si}{14} Ly$\alpha$, \ion{Ne}{10} Ly$\beta$, \ion{N}{7} RRC,
and \ion{O}{8} RRC.
Collectively, these features reveal the presence of a photoionized plasma,
which also produces absorption edges of \ion{O}{7} and \ion{Mg}{11}
in the spectra. 
The fluorescence lines
(Si K$\alpha$, S K$\alpha$,  and Fe K$\alpha$)
are not detected at the 3$\sigma$ level. 
The recombination lines have larger EWs 
during dips, because the X-ray continuum 
is weaker during the dip states than during the persistent states. 

In the following, we constrain the density, 
temperature, and dynamics of the photoionized plasma.
We measure the elemental abundance ratios in the emitting plasma. 
We also test continuum models
for the dip and persistent state spectra, and we identify
the ionization level of the absorber.

\subsection{Diagnostics with Helium-like ion lines}
\label{sub:helike}

\begin{figure}
\epsscale{.5}
\plotone{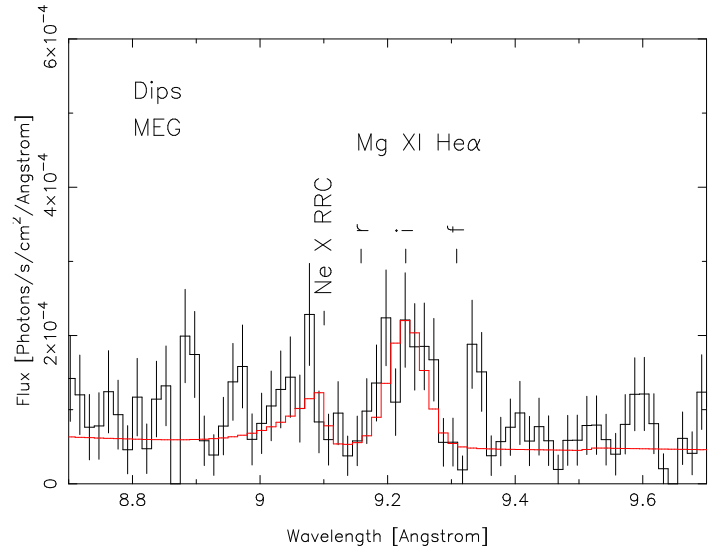}
\caption{Observed MEG spectrum of the \ion{Mg}{11} He$\alpha$ line region during dip states (black)
and model fit (red). The $G = (f+i)/r$ line ratio
implies a photoionized plasma. The high $R=i/f$ line ratio is due
to either a high electron density or a strong UV field near the accretion disk
(\S \ref{sub:helike}).
\label{fig:mgxi}}
\end{figure}

\begin{figure}
\epsscale{.5}
\plotone{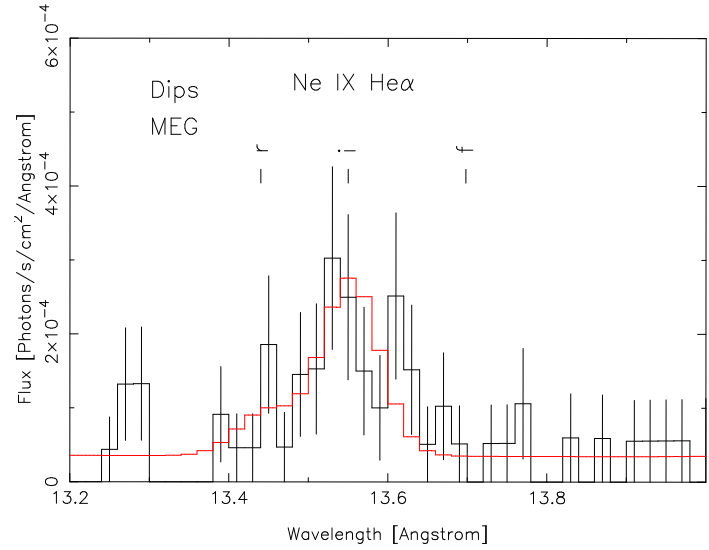}
\caption{Observed MEG spectrum of the \ion{Ne}{9} He$\alpha$ line region during dip states (black) and
model fit (red).
\label{fig:neix}}
\end{figure}

\begin{figure}
\epsscale{.5}
\plotone{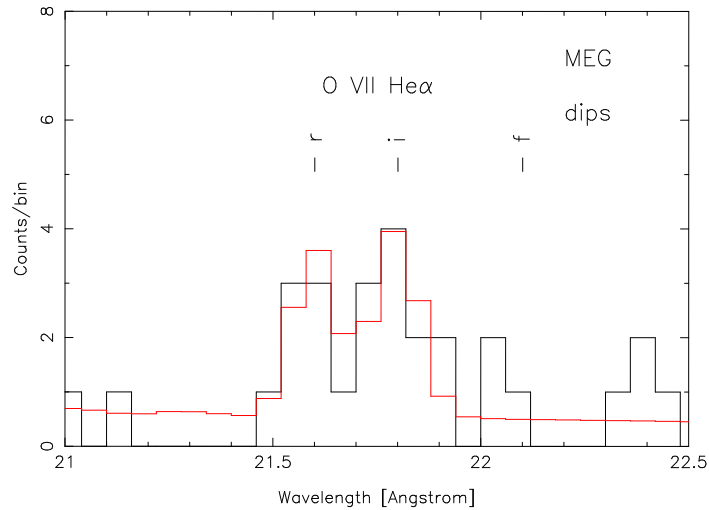}
\caption{Observed MEG spectrum of the \ion{O}{7} He$\alpha$ line region during dip states (black), which shows that
\ion{O}{7} He$\alpha$ is detected. The model fit is shown in red.
\label{fig:ovii}}
\end{figure}

We detect emission from three He-like ions: \ion{O}{7}, \ion{Ne}{9},
and \ion{Mg}{11}. The He$\alpha$ lines are shown in Figures
\ref{fig:mgxi}, \ref{fig:neix}, and \ref{fig:ovii}.
The He$\alpha$ line ratios can be used to determine the
dominant heating mechanism in the plasma \cite[]{liedahl99}.
The He$\alpha$ line complex can also be used to constrain the electron density 
($n_e$) and/or the ambient UV flux \cite[]{gabriel,blum}.
Three lines in the He$\alpha$ complex 
are resolvable with the HETGS: the resonance line 
($r$ denotes $2^1P_1 \to 1^1S_0$), the
intercombination line ($i$ denotes the two blended transitions
$2^3P_{1,2} \to \ 1^1S_0$), 
and the forbidden line ($f$ denotes $2^3S_1 \to 1^1S_0$).
The notation $n^{2S+1}L_{J}$ identifies each two-electron atomic state,
with quantum numbers $n$, $L$, $S$, and $J$.
Weak satellite lines are also present in the He$\alpha$ complex.

In \exop, the $i$ line dominates all the He$\alpha$ complexes.
The $r$ and $f$ lines are not detected, perhaps
with the exception of the $r$ line
of \ion{O}{7}, for which a few photons are detected.
We now discuss the consequences 
of the observed He$\alpha$ line ratios.

The plasma must be heated by photoionization.
In a photoionized plasma, recombination is usually followed by
RRC emission, and the subsequent bound electron cascades
produce the line emission. This contrasts with collisionally ionized gases,
for which line emission is predominantly produced after
resonant excitation by electron impact.
This difference is measurable by the line ratio $G=(f+i)/r$, which is
$G \sim 4$ for photoionized gas and   
$G \lesssim 1$ for collisionally ionized gas \cite[]{liedahl99,porquet}. 
The $G$ ratios of \ion{O}{7} and \ion{Ne}{9} measured
with \xmm RGS were interpreted by C01 as a signature of a recombining plasma.
The \chandra HETGS data in Table \ref{tab:helike} show that the
$G$ ratios of \ion{O}{7} and \ion{Ne}{9} are consistent with the C01 measurements,
and similarly, that $G \gtrsim 3$ for \ion{Mg}{11}. These $G$ ratios 
indicate that photoionization is the dominant heating mechanism for a wide range
of ionization parameters.

The plasma density is very high, or the
UV field is very intense near the He$\alpha$ emission region, or both.
This is based on the upper limits we set on the $R=f/i$ line ratio
from helium-like ions (Table \ref{tab:helike}).
The $R$ ratio of the plasma depends on the effects of electron-impact excitation
at high density and/or photoexcitation by an intense UV field
\cite[]{gabriel,blum,photoex}.
In the limit of a weak UV field ($F_\lambda \ll F_\lambda^{\rm crit}$),
the density $n_e$ is constrained to be
above a critical value $n_e > n_e^{\rm crit}$.
Conversely, in
the low density limit ($n_e \ll n_{\rm crit}$),
the net UV flux $F_\lambda$
is constrained to be above a critical value $F_\lambda > F_\lambda^{\rm crit}$
at $\lambda = 1637$~\AA, 1270~\AA, and 1033~\AA, which correspond
to the $2^3S_1 \to 2^3P_{0,1,2}$ (or $f \to i$) transitions of
\ion{O}{7}, \ion{Ne}{9}, and 
\ion{Mg}{11}, respectively. 
Note $F_\lambda$ is the UV flux local to the He$\alpha$ 
emission region in units of \ergpcmsqpspA.
We conservatively estimate $F_\lambda^{\rm crit}$ by
equating $w_f = w_{f \to i}$, where $w_f$ is the
$2^3S_1 \to 1^1S_0$ 
decay rate 
from \cite{drake}, and
$w_{f \to i}$ is the
$2^3S_1 \to 2^3P_{0,1,2}$
photoexcitation rate. \cite{kahn} used a similar procedure
to set limits on the distance from the He$\alpha$ emission region to the UV source.
The photoexcitation rate is
\begin{equation}
w_{f \to i} = \frac{\pi e^2 \lambda^3 F_\lambda f_{\rm osc}}{h m_e c^3},
\end{equation}
where $e$, $m_e$ are the electron charge and mass, $c$ is the speed of light,
and $h$ is Planck's constant.
We calculate $w_{f \to i}$ 
by using the oscillator strengths $f_{\rm osc}$ for
the $2^3S_1 \to 2^3P_{0,1,2}$
transitions in \ion{O}{7} and \ion{Ne}{11} from \cite{cann}
and in \ion{Mg}{11} from Duane Liedahl (private communication, 2002).
C01 attributed the $R \sim 0$ ratio in \ion{O}{7} and \ion{Ne}{9}
observed in \exo to electron-impact 
excitation, thereby setting lower limits to the density. 
We set both density and UV flux limits, as shown in Table \ref{tab:helike}. 
Photoexcitation may be important for
\ion{N}{6} through \ion{Mg}{11}, and the limits we set on the
UV field make up for the absence of UV spectral data.
A detailed calculation of the electron populations
in the $n=2$ shell of He-like ions
in photoionization equilibrium is needed
to determine accurate limits on the space of
allowed values of $(F_\lambda, n_e)$.

In \exop, the electron temperature is low compared to that of a collisionally
excited gas of equivalent ionization. The $G$ ratio of \ion{Mg}{11} implies an electron
temperature $T < 3 \times 10^6$~K. Similarly, for \ion{Ne}{9}
we get $T < 2 \times 10^6$~K and for \ion{O}{7} 
we get $T < 1 \times 10^6$~K. 
In \S \ref{sub:temp}, we improve upon the 
above $T$ limits by use of the RRC. 

\subsection{Temperature diagnostics}
\label{sub:temp}

We detect \ion{Ne}{9} RRC and \ion{Ne}{10} RRC, at a
3$\sigma$ and 4$\sigma$ level respectively,
which confirms that the plasma is photoionized and allows us to constrain
$T$.  Free electrons recombining with ions produce an RRC 
width proportional to $T$ \cite[]{liedahlrrc}.
In \exop, the velocity broadening observed in the
recombination lines implies that
for the RRC, velocity and temperature broadening 
are of the same order.
We fit the RRC profiles neglecting velocity broadening,
and the resulting 3$\sigma$ upper limit is
$kT \lesssim 20$~eV or $T \lesssim 10^5$~K for these ions.
This $T$ is in the expected range calculated using
the XSTAR \cite[]{kallman82} photoionized plasma equilibrium model.

\subsection{Line profiles}
\label{sub:profiles}
The measured velocity broadening
of the three brightest lines
is consistent with a constant value of $\bar{\sigma}_v \sim 750 \pm 120$\kmps
(see Table \ref{tab:lines}). 
\exo and 4U~1626-67 \cite[]{4u1626} are the only
LMXB known to exhibit broadening in their
soft X-ray emission lines, which are likely indicative of Kepler motion 
in the accretion disk. 
We fit the lines with Gaussian profiles using the \cite{cash} statistic, 
which is appropriate for Poisson errors.
The statistical error on the velocity broadening
is $\sigma_v /\sqrt{N}$, where $N$ is the number of photons
in the line. Due to the small number of counts, 
the remaining lines are not included in the computation
of $\bar{\sigma}_v$ above, and instead we fix
their $\sigma_v$. This is done since the Poisson
fluctuations in the continuum
dominate the profiles of the weakest lines. 

The HETGS spectra show that
the velocity broadenings ($\sigma_v$) of 
the lines from one ionic species to another
are equal within $\pm 330$\kmps. 
Observations of \ion{N}{7}, \ion{O}{8}, and \ion{Ne}{10}
with the \xmm RGS revealed a trend
in $\sigma_v$ with ionization state, with differences as large
as $1750 \pm 500$ \kmps (C01).
Our HETGS data for \ion{Ne}{9}, \ion{Ne}{10}, and
\ion{Mg}{11}, show smaller line widths than those
measured with RGS and smaller differences in $\sigma_v$ from one line to another
(Table \ref{tab:lines}).

\subsection{An absorbing medium composed of ionized and neutral
gas}
\label{sub:edges}

Our fits of the HETGS spectra show that dips can be produced by 
an absorbing medium composed of both neutral and
ionized gas (or ``warm absorber''). The medium selectively absorbs 
the hard X-ray continuum, but not the recombination emission,
which is only absorbed by interstellar gas. We assume
Solar abundances for the neutral component of the absorber.

\begin{figure}
\epsscale{0.7}
\plotone{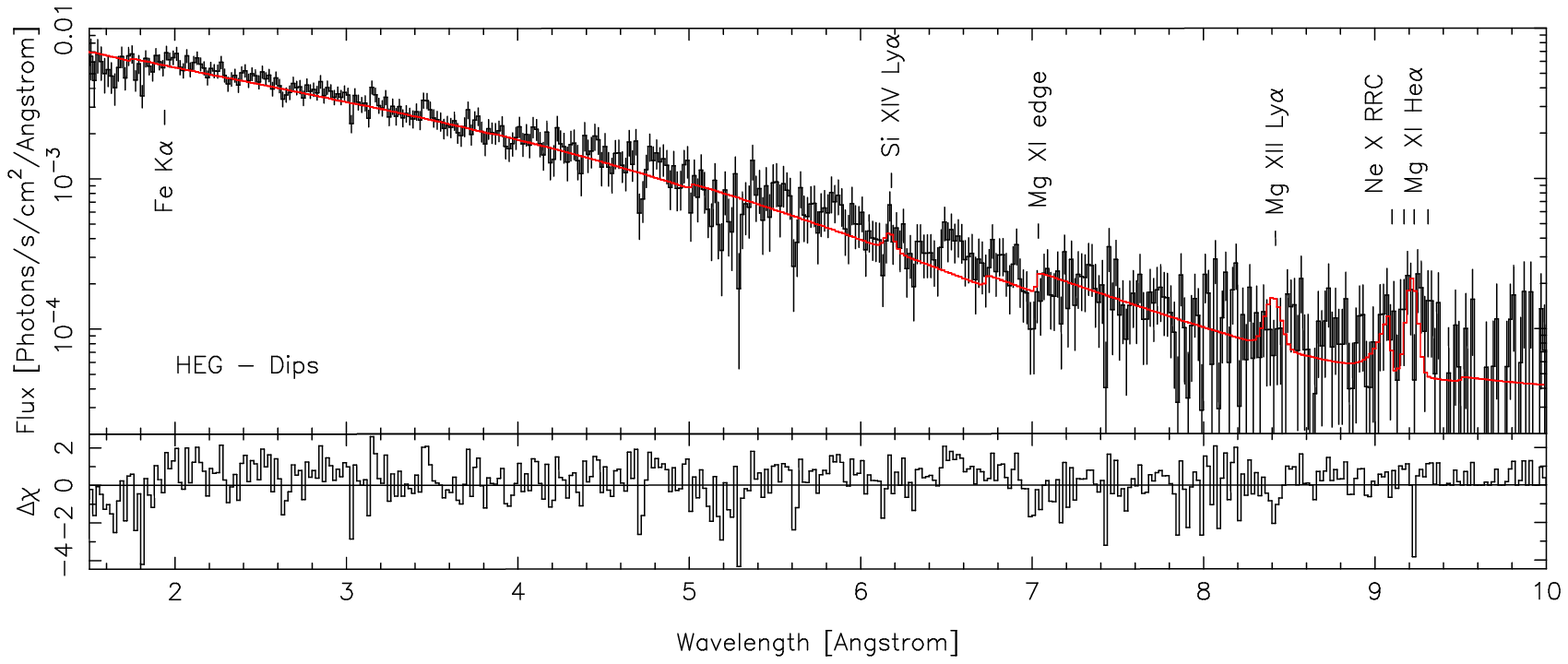}
\plotone{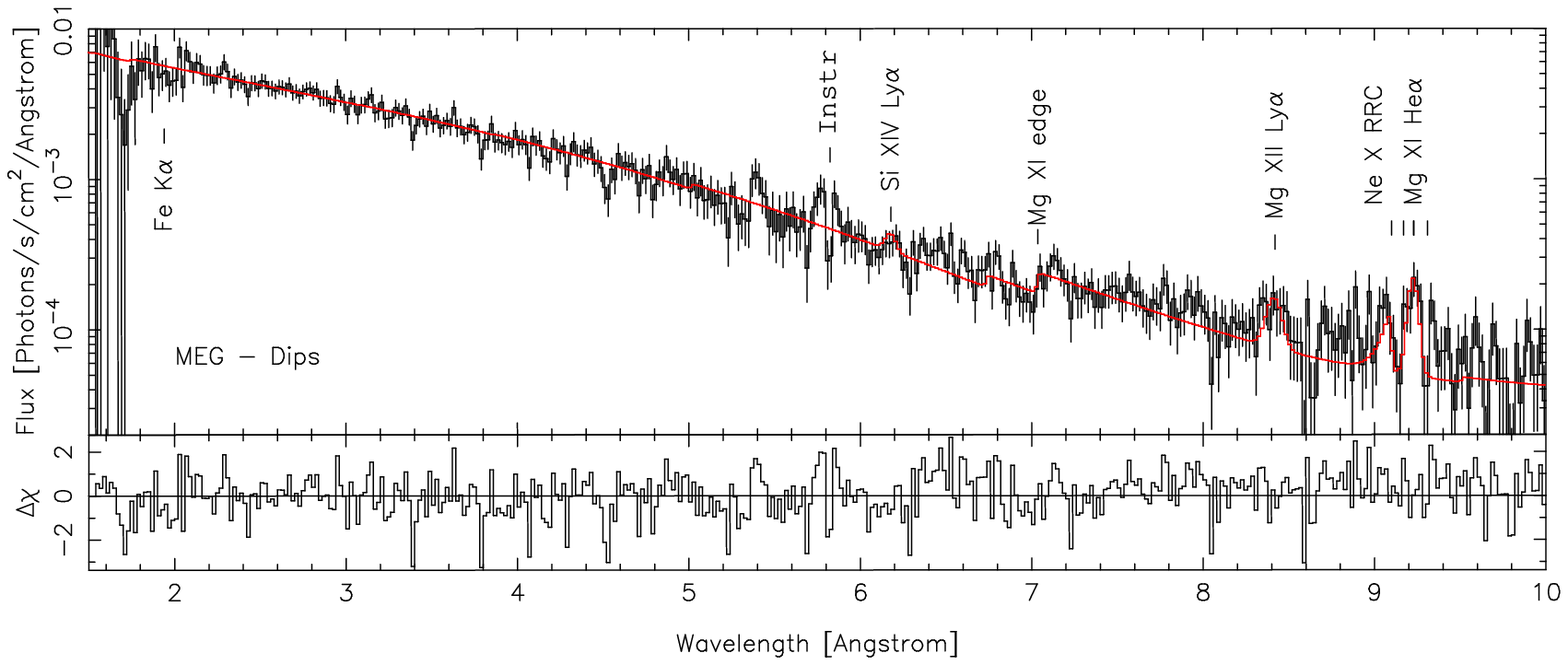}
\caption{Observed spectra during the 29.5~ks dip state (black), with model fit (red),
and residuals (bottom, in units of the standard deviation). 
Grating: (a) HEG, (b) MEG. The \ion{Mg}{11} $i$ line and the \ion{Mg}{11} K edge
are most prominent in the MEG spectrum.
The discrete spectral features, which include RRC, are the signatures of a
photoionized plasma.
\label{fig:dip}}
\end{figure}

\begin{figure}
\plotone{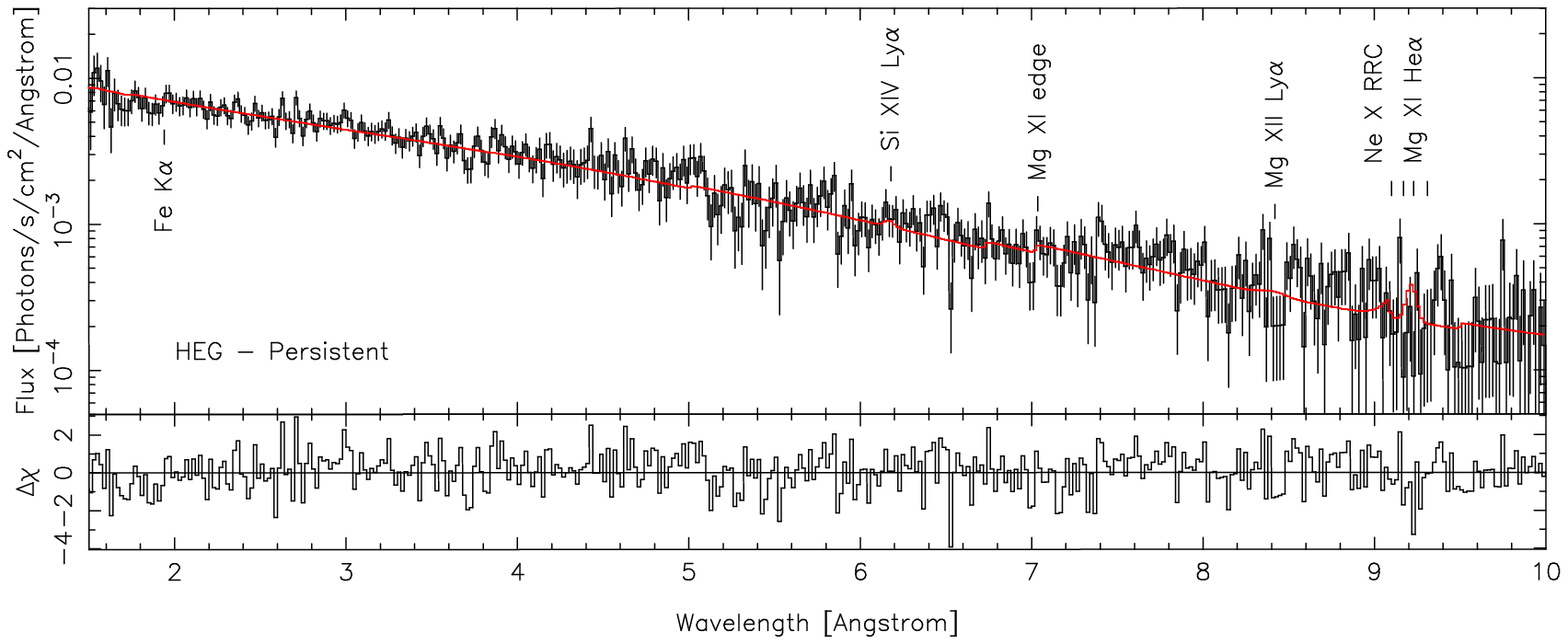}
\plotone{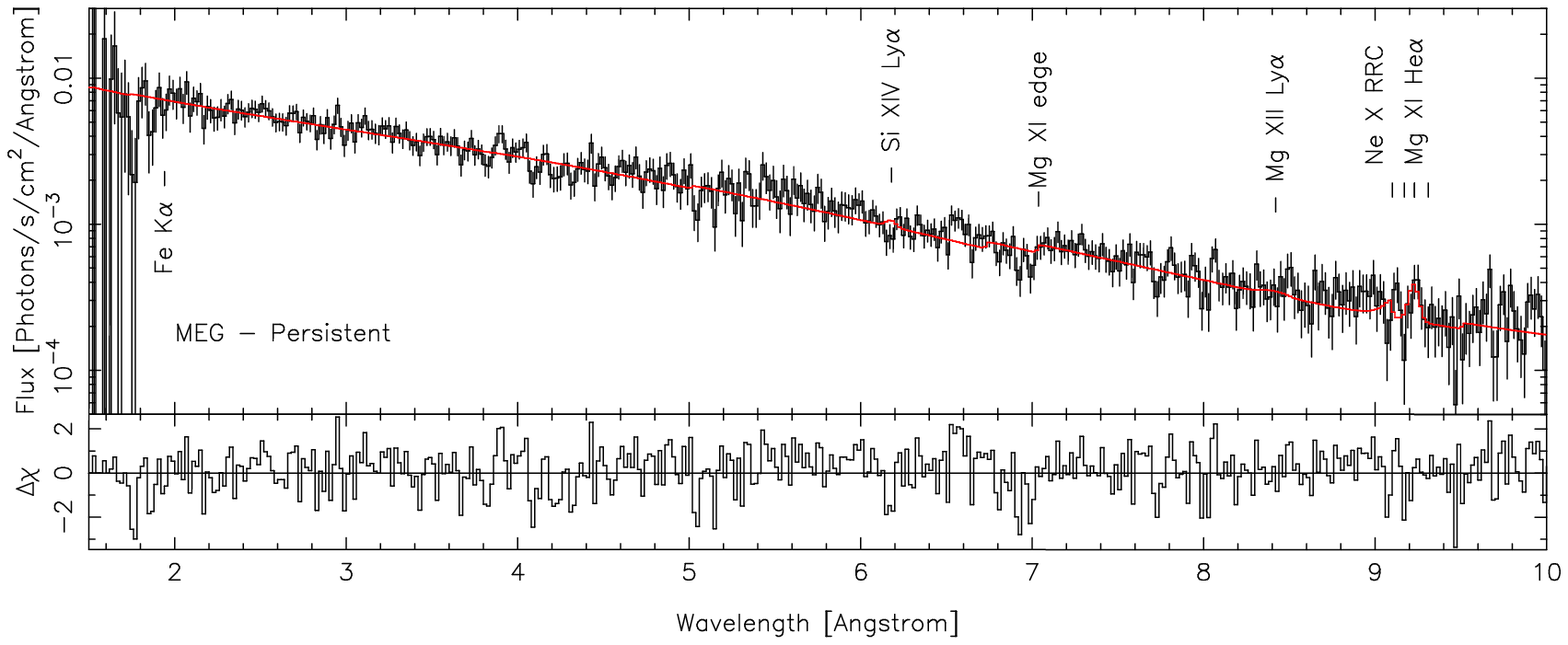}
\caption{Observed spectra during the 15.7~ks persistent state (black), with model fit (red),
and residuals (bottom, in units of the standard deviation). 
Grating: (a) HEG, (b) MEG. Here, the discrete spectral features are not as prominent as
in Fig. \ref{fig:dip}.
\label{fig:pers}}
\end{figure}

\begin{figure}
\epsscale{1}
\plotone{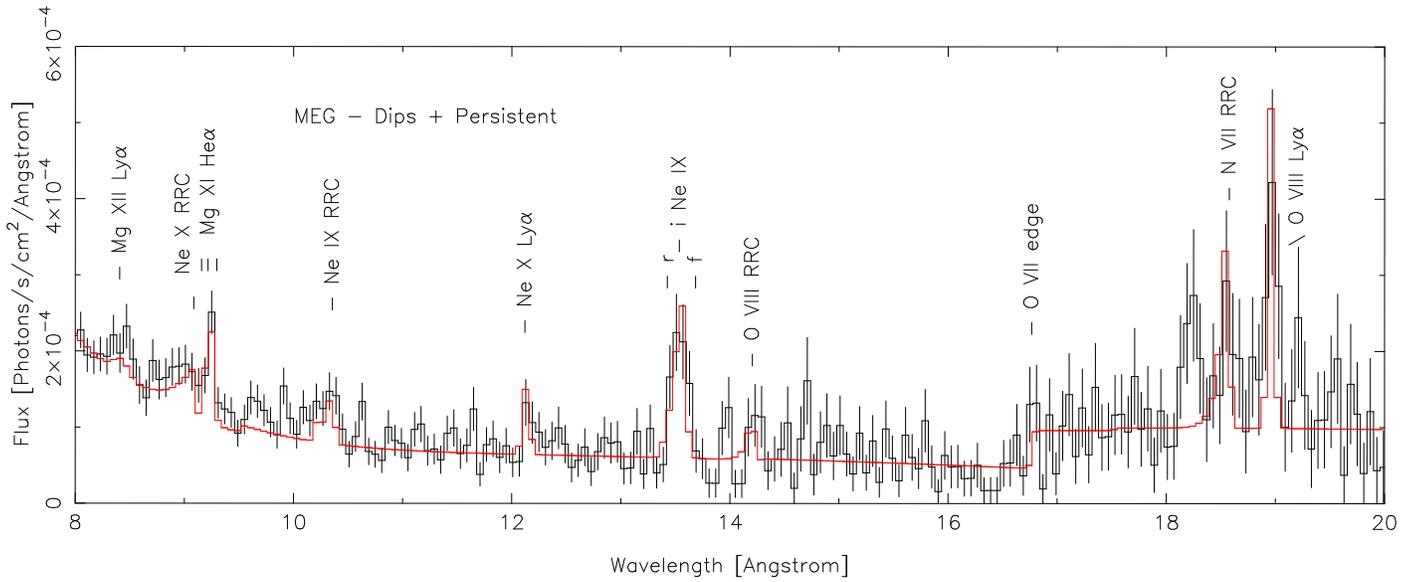}
\caption{Observed MEG spectrum including both the dip and persistent states (black) and model fit (red).
Bursts are excluded. The line broadening, especially
discernible for \ion{Ne}{9} $i$, and the limits on the systemic velocity,
can be explained if the plasma is orbiting the neutron star.
The \ion{O}{7} K shell absorption edge shows that part of the material
producing the intensity dips is highly ionized. 
\label{fig:tot_detail}}
\end{figure}

We detect both \ion{O}{7} K and \ion{Mg}{11} K absorption edges
(the edge optical depths are shown in Table \ref{tab:edges}).  
The \ion{O}{7} K edge is detected only during
the persistent state, and it implies
$N_{\rm O VII} = (3.2 \pm 0.6) \times 10^{18}$\pcmsq \ (Fig. \ref{fig:tot_detail}).
We also find a \ion{Mg}{11} K edge (see Figs. 
\ref{fig:dip} and \ref{fig:pers}). The \ion{Mg}{11} K edge is 
most prominent in the MEG grating order which
is {\it not} affected by a chip gap. 
The \ion{Mg}{11} column density is constant or slightly increasing
from the persistent to the dip state, 
from $N_{\rm Mg XI} = ( 1.6 \pm 0.6 ) \times 10^{18}$\pcmsq \
to $N_{\rm Mg XI} = ( 3.6 \pm 0.7 ) \times 10^{18}$\pcmsq.
For comparison, the \nH absorber on the hard continuum increases by a factor of two 
from the persistent to the dip state
(see Tables \ref{tab:contpl} and \ref{tab:contpc}).  

\subsection{Elemental abundance ratios derived from recombination emission}
\label{sub:abun}

We use the method introduced by \cite{jimenez} to measure
the elemental abundance ratios. In this method, we
use a recombination plasma model with a finely spaced distribution
of ionization parameters 
to fit the observed emission.
We calculate a grid of thermal equilibrium 
models versus ionization parameter $\xi$, using the
XSTAR plasma code \cite[]{kallman82}.
We fit the differential emission measure as a function of $\xi$
with a power-law distribution of the form
\begin{equation}
\label{eq:dem}
{\rm DEM}(\xi) = \frac{\partial({\rm EM})}{\partial(\log_{10} \xi)} = K \xi^{\gamma} ,
\end{equation}
where $\xi= L /( n_e r^2 )$,
$K$ and $\gamma$ are fit parameters, 
 and the emission measure is
${\rm EM} = \int n_e^2 dV$.
We use the recombination
rate coefficients calculated 
by Liedahl (private communication)
with the HULLAC code \cite[]{hullac}, to
calculate the emission line fluxes.
We fit the spectral data during the dip state, where the
H-like and He-like lines of O, Ne and Mg are detected.
We fit six spectral lines with four parameters:
two emission measure parameters plus two abundance ratios.
Thus, the fit is over-constrained.
Our results are shown in Table \ref{tab:elem}.
We find that the O/Ne abundance ratio is smaller than the
Solar value by 3$\sigma$. The Mg/Ne ratio is within
2$\sigma$ of the Solar value. We caution that the
detection of more ionic species is desirable to further constrain 
the DEM and to make the Mg/Ne abundance measurements more robust.

Our results are consistent with the O/Ne abundance
ratio estimated by C01, who defined an 
ionization parameter of formation to
estimate the emission measure. They also estimated
a Solar value for the N/Ne abundance ratio.
We assign a Solar abundance to Ne.

\subsection{Continuum emission} 
\label{sub:cont}
We test partial covering and two-power-law continuum model
fits of the HETGS spectra.
The parameters of continuum fits 
using a two power-law model are
shown in Table \ref{tab:contpl}.
The parameters of the fits using a partial covering
model are shown in Table \ref{tab:contpc}.
The fit results for both the dip and persistent states are shown.
We find a very weak soft component with both models
(fits without it are just marginally worse).
The goodness-of-fit from the Cash statistic
is the same for either model. 

The partial covering model requires
a change in the normalization of the hard
power-law. It is not able to
fit the data well with just a change in the covering fraction and
\nH. The intensity changes by $\sim 10$\%
from the dip to the persistent state, which
may be due to a change in the accretion rate.
The model fits with the two power-law model
for the dip state are shown in Figure \ref{fig:dip}
and for the persistent state in Figure \ref{fig:pers}. 
In the $1.6 < \lambda < 6.9$~\AA \ range, we do not find
a statistically significant 
spectral feature, following the criteria in \S \ref{sec:obs}.

\section{Discussion}
\label{sec:disc}

The recombination emission and the \ion{O}{7} and \ion{Mg}{11} K-shell absorption
edges observed in \exo constrain the density and location of
the photoionized plasma, which appears coincident with an outer
disk thickened by X-ray radiation and by 
stream-disk impact. The thickened disk is not necessarily out of
hydrostatic equilibrium.
Aside from eclipses and type I bursts,
we attribute the continuum variation to
dipping due to absorption from a mix of ionized and neutral gas
in the line of sight. The absorbing material is likely in the outer disk. 
The observed He-like ion emission lines imply that the plasma 
is at high density or there is a strong UV
field, either of which provide evidence for
the proximity of the line emission region to the accretion disk.
We set a limit on the Mg/O abundance ratio implied by the absorption edges and
compare the result to the ratio derived from the emission lines.

\subsection{The structure and density of the absorber and the soft component}

The absorber, composed of neutral and ionized gas,
is located at the outer accretion disk, at
a height $0.14r < h < 0.27r$ above the disk midplane,
which is determined from the inclination $i$ (see \S \ref{sec:intro}).
Our spectral fits do not exclude the possibility that 
the addition of other ionized species can fully account for the absorption.
The periodic dips are produced by gas just inside the 
Roche Lobe $R_{\rm L}$ of the neutron star.
The dips at orbital phase $\phi \sim 0.9$ are produced by gas
located at the point where the accretion stream impacts the disk. 
The dips at $\phi \sim 0.6$ may originate
in a warp or bulge on the outer rim of the disk.
The stochastic dips are either due to stochastic
structure at the outer disk rim or are 
located at the inner disk. 
Whereas C01 reported no measured change in the 
K-shell absorption edges of \ion{O}{7} and \ion{O}{8},
we find evidence for variability of the optical depth of 
the K-shell \ion{Mg}{11} edge, which is correlated with dip activity.
The spectral changes 
imply that all the intensity variations outside eclipses
and bursts are due to dips, and that these dips are 
produced by column density variations of the partially ionized 
absorber.

The soft X-ray continuum region is spatially compact, whereas
the recombination emission region is extended.
The emission region size can be inferred 
from the orbital phase variations in the light curve.
C01 concluded that both continuum and line emission regions
should be extended, because the 6--35 \AA \ intensity
was independent of orbital phase.  
An extended line region is consistent with our HETGS results, 
since the line fluxes are not affected by dips 
(the region size is measured via the line widths in \S \ref{sub:disk}).
However, from the eclipse egress duration of $\lesssim 50$~s
in the 6--22 \AA \ band, we constrain the soft X-ray continuum
region size to $\lesssim 3 \times 10^9$\cm. 
The eclipses observed with the {\it Rossi X-ray Timing Explorer}
yielded a source size of (1--3)$ \times 10^8$\cm \ in the 0.6--6.2 \AA \ band
(J. A. Jenkins \& J. E. Grindlay, in preparation).
A relative orbital speed of $v \sim 500$\kmps was used in both
measurements.

The ionized absorption and X-ray recombination emission can originate
in the same region, as suggested by C01. 
From the derived $N_{\rm Mg XI}$ (\S \ref{sub:edges}),
we find $n_e > 7 \times 10^{11} ( {\rm Mg}_\sun / {\rm Mg} )$ \pcmcu
for the persistent state along the line of sight, and 
$n_e > 2 \times 10^{12} ( {\rm Mg}_\sun / {\rm Mg} )$ \pcmcu
for the dip states, since 
the absorption region has $r < R_{\rm L}$
(${\rm X}_\sun$ and ${\rm X}$ denote the Solar and observed
abundances for element X, respectively).
These density limits agree with those
derived for the \ion{Mg}{11} in emission (Fig. \ref{fig:mgdiag}),
and therefore the same region can produce the observed
absorption and emission.
Similarly with $N_{\rm O VII}$, we set 
$n_e > 6 \times 10^{10} ( {\rm O}_\sun / {\rm O} )$\pcmcu, which
is consistent with the R ratio.

Our finding that the absorber has two components with 
distinct ionization parameters supports the picture proposed by
\cite{incl} to explain the origin of the intensity dips in LMXBs. 
In this picture, the presence of episodic dips is explained by
the transit of clouds embedded in a two-phase medium
which is produced by photoionization. 

\subsection{Constraints on accretion disk properties}
\label{sub:disk}

\begin{figure}
\epsscale{1.}
\plotone{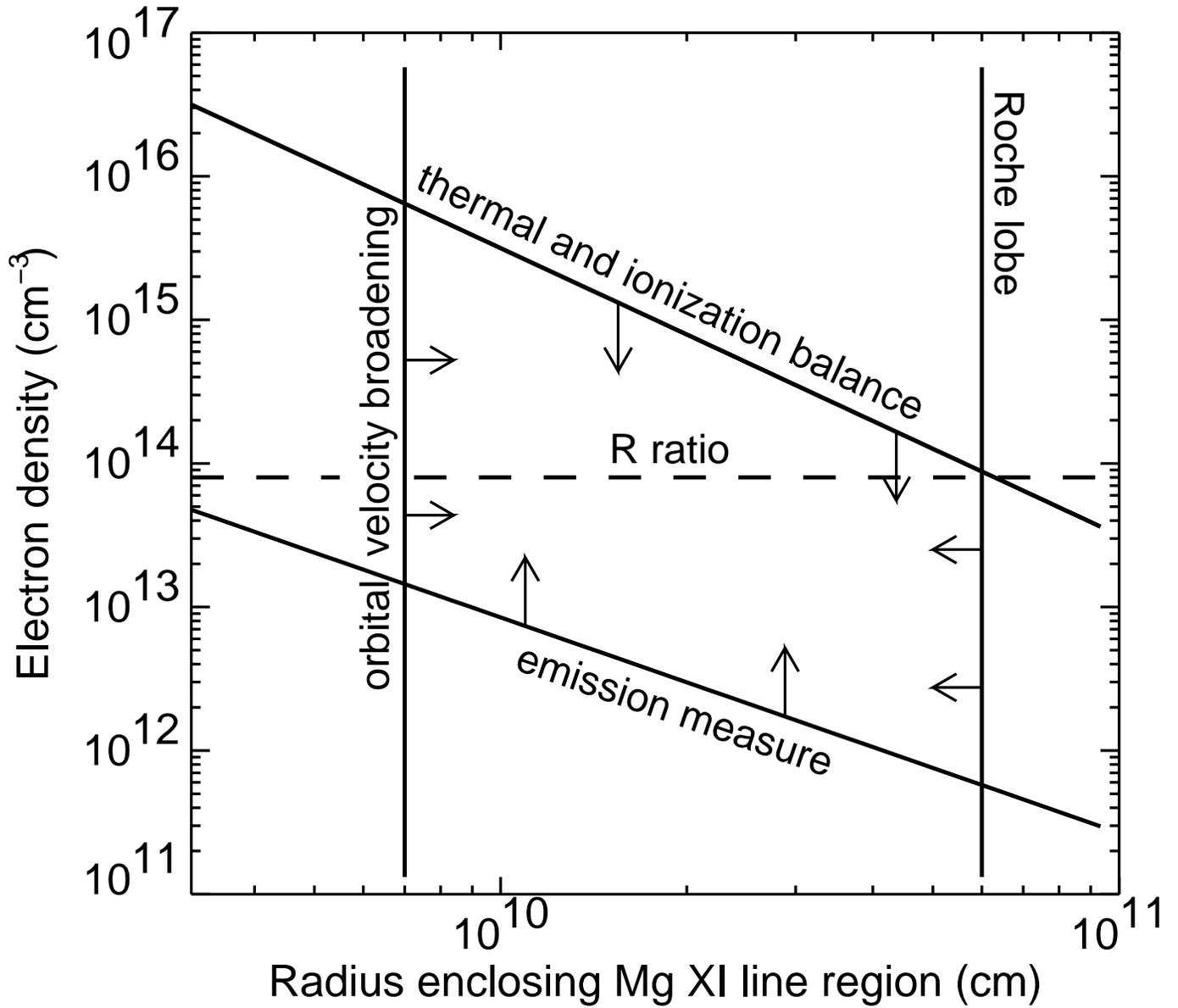}
\caption{Derived limits on the region which produces the \ion{Mg}{11} He$\alpha$ recombination emission.
The radius is the distance from the neutron star. 
The \ion{Mg}{11} region is located near the outer accretion disk.
\label{fig:mgdiag}}
\end{figure}

The spectroscopic evidence points to a recombination region which is located
above the outer accretion disk.  We constrain the 
density and location of the 
\ion{Mg}{11}
emission region,
based on
1) the ionization level of \ion{Mg}{11} calculated with photoionization
equilibrium models, 2) the emission measure derived from
the \ion{Mg}{11} He$\alpha$ flux, and 3) the interpretation of the
\ion{Mg}{11} velocity broadening as circular Keplerian motion
around the neutron star. 
We set $n_e < L_{\rm max}/ \xi_{\min} r^2$,
where $\xi_{\min}$ is defined such that
for $\xi > \xi_{\min}$,
$\sim 90$~\% of the \ion{Mg}{11} line flux is emitted, 
and $L_{\rm max} = 10^{37}$\ergps is the unabsorbed X-ray luminosity.
The model in \S \ref{sub:abun} yields
$\log_{10} \xi_{\min} \simeq 1.5$ ($\xi$ is in units of \ergcmps).
To set a lower limit on the rms value of $n_e$, we obtain
the EM of the \ion{Mg}{11}-emitting
plasma from the $K \xi^\gamma$ fit in Table \ref{tab:elem}.
We derive upper limits for
the systemic velocity of $v \lesssim 310$\kmps for
\ion{Ne}{9} He$\alpha$, $v \lesssim 470$\kmps for
\ion{Mg}{11} He$\alpha$, and $v \lesssim 790$\kmps for
\ion{Mg}{12} Ly$\alpha$ (at 90\% confidence; 
for the He$\alpha$ lines, we keep $G$ and $R$ fixed).
Such small systemic velocities 
are expected if the emitting plasma is 
orbiting the neutron star (C01). 
The EWs of the double-peaked optical emission lines 
with ${\rm FWHM} \sim 2000$\kmps are 
enhanced during X-ray eclipses, which 
were interpreted as emission from a disk with 
$r \sim 6 \times 10^{10}$ cm \cite[]{optical_curves_spec}.
For circular Kepler orbits, and in the absence of optical depth effects,
$\sim 90$\% of the line emission
occurs at $r \gtrsim G M_{NS} / (2 \sigma_v \sin i)^2$. This $r$ limit
is less dependent on the disk emissivity $\epsilon(r)$ than 
the characteristic $r$ derived by C01.
The set of $r$ and $n_e$ limits 
in Figure \ref{fig:mgdiag} show that
the \ion{Mg}{11} emission region is fully
consistent with a plasma in orbit
inside the primary's Roche lobe.
The threshold density $n_e^{\rm crit}$ derived from
the \ion{Mg}{11} $R$ ratio in \S \ref{sub:helike} bisects the allowed $n_e$
range.  

Our measurements show that the 
line emission originates predominantly at the outer decade in radius.
The $\sigma_v$ we measured for \ion{Mg}{11}, \ion{Ne}{9}, and \ion{Ne}{10} 
are in agreement with the C01 measurements
for \ion{N}{7}. The $\sigma_v$ for \ion{Ne}{10} and \ion{O}{8}
reported by C01 are significantly larger than our values. 
Whereas C01 found a correlation between $\sigma_v$
and the ionization level of the emitting ion, 
we find no evidence for such a correlation (see \S \ref{sub:profiles}).
This implies that the emissivities of different ions have
similar $r$ dependence. 
In that case, for ions at a given $r$ from the neutron star,
a density gradient is needed to produce distinct ionization parameters.
Instead of the radially layered ionization structure 
proposed by C01, we find that vertical stratification of the disk density 
at each radius can explain the observed line widths.

The presence of photoionized plasma 
at large disk radii and at a height $h \gtrsim 0.14~r$ 
requires a mechanism to expand the disk in the vertical direction. 
The disk temperature predicted by the \cite{ss73} model 
for $r \gg R_{NS}$ is
\begin{equation}
\label{eq:sstemp}
T = 1.4 \times 10^{4} ~ \dot{M}_{17}^{1/4} m_{1.4}^{-1/4} r_{10}^{-3/4}~{\rm K},
\end{equation}
and it is too low to explain the presence of the H-like and He-like ions, where
$\dot{M}_{17}$ is the accretion rate in units of $10^{17}$\gps, 
$m_{1.4}$ is the neutron star mass in units of 1.4~$M_\sun$, 
and $r_{10}$ is the radius in units of $10^{10}$~cm.
For $r \gg R_{NS}$, the \cite{ss73} scale height to radius ratio is
\begin{equation}
h/r = 2.1 \times 10^{-2} ~ \alpha^{-1/10} \dot{M}_{17}^{3/20} m_{1.4}^{-3/8} r_{10}^{1/8} ,
\end{equation}
which is too small to explain the absorption edges at $h/r \gtrsim 0.14$.
C01 suggested through a similar argument that the disk material 
must be far from hydrostatic equilibrium in the vertical direction. 

We propose that two mechanisms cause
the disk expansion: 
1) the impact of the accretion stream with the disk; and
2) the illumination from the neutron star,
since the disk is heated and its scale height increases.
Firstly, the ram pressure at the site of the impact of the accretion stream
with the outer disk will increase the disk thickness. The increase
in disk thickness depends on the ratio of the ram pressure to the outer disk
pressure, as well as on the cooling timescale of the gas compared
to the dynamical timescale. Theoretical calculations show that 
at $i = 78$\degr, the pre-eclipse dips in \exo can be understood as a result
of absorption from a bulge produced by disk-stream impact \cite[and references therein]{arlivio}. The mid-orbit
dips could be due to a similar impact point at smaller radii.
Secondly, as shown in Figure \ref{fig:mgdiag}, photoionization equilibrium
models show that irradiation from the observed neutron star continuum 
can energize the disk and produce the line emission. To illustrate this,
consider that the effective temperature of the neutron star continuum is
\begin{equation}
T_{\rm eff}  = 1.1 \times 10^5 ~ L_{37}^{1/4} r_{10}^{-1/2}~{\rm K},
\end{equation}
where $L_{37}$ is the luminosity in units of $10^{37}$\ergps. 
Another effect of photoionization heating is to increase the disk 
photosphere scale height significantly at large radii by roughly
\begin{equation}
h_{\rm phot}/h \sim (T_{\rm eff}/T)^{1/2} \sim 3 ~ \dot{M}_{17}^{-1/8} m_{1.4}^{1/8} L_{37}^{1/8} r_{10}^{1/8} .
\end{equation}
Models of a centrally illuminated accretion disk atmosphere and corona, which 
include temperature and pressure gradients as well as radiation transfer,
yield atmospheric scale heights $h/r \sim 0.11$ which are compatible with
the observed ionized absorption, with model parameters $L = 10^{37.3}$\ergps and
$r = 10^{10.8}$\cm \ 
\cite[]{diskm,liedahlme}. These models show that an atmosphere heated
by irradiation can produce a geometrically thickened disk in hydrostatic 
equilibrium. 

While a detailed comparison of the data with accretion disk atmosphere models
is outside the scope of this article, the models are constrained
by the measured line fluxes, the line profiles, and by the
DEM($\xi$). For example, the behavior of the
line profiles for ions at different ionization levels provides information
on the disk structure. A disk in which the DEM is set by vertical stratification 
will yield nearly equal $\sigma_v$ for all ions, while 
if the DEM is set by radial stratification,  
$\sigma_v$ will increase with $\xi$.

\subsection{UV emission from the disk can photoexcite He-like ions}

We use the observed $R$ ratio to measure 
the UV emission of the accretion disk, for
the low density case.
Having established the location of the
recombination region by other means, we can use
the $R$ ratios to estimate the UV luminosity of the disk.
We use the limits on $F_\lambda$ 
at $\lambda = 1033$~\AA, 1270~\AA, and 1637~\AA, which are local to
the recombination region. The limits 
from photoexcitation calculations assume $n_e \ll n_e^{\rm crit}$ and
are shown in Table \ref{tab:helike} (see also \S \ref{sub:helike}).
Taking a fiducial
distance between the recombination region
and the disk photosphere to be
$h \sim r / \tan i \sim 10^{10}$~cm, we get
$L(1033$\AA$) > 2 \times 10^{29}~h_{10}^2$\ergpspa,
$L(1270$\AA$) > 2 \times 10^{28}~h_{10}^2$\ergpspa, and
$L(1637$\AA$) > 6 \times 10^{26}~h_{10}^2$\ergpspa. 
The UV emission of a model disk
with $L = 10^{37.3}$\ergps
is calculated using equation (\ref{eq:sstemp})
plus a neutron star illumination term, with
the disk albedos calculated by \cite{diskm}.
The model disk luminosities (for $r \sim 10^{11}$~cm) are 
$L(1033$\AA$) \sim 7 \times 10^{31}$\ergpspa,
$L(1270$\AA$) \sim 5 \times 10^{31}$\ergpspa, and
$L(1637$\AA$) \sim 4 \times 10^{31}$\ergpspa.
If $r \sim 3 \times 10^{10}$~cm, 
$L(1637$\AA$)$ is $\sim 20$\% lower
and $L(\lambda)$ is unchanged for
$\lambda \lesssim 1400$~\AA.
We conclude that in the low density limit,
the derived lower limits on the
UV luminosity are fully consistent with
the expected disk emission.

The effects of UV photoexcitation of $2^3S_1 \to 2^3P_{0,1,2}$
in the $R$ ratio of low- and mid-$Z$ He-like ions were identified 
in the spectra of stellar coronae and X-ray binaries.
Photoexcitation was taken into account to accurately measure the density
in the Solar corona with \ion{C}{5} \cite[]{gabriel,blum,suncv}.
Photoexcitation by the UV field in the vicinity of O-stars 
was found to overwhelm the effect of electron-impact 
excitation, producing $R \lesssim 0.3$ ratios of \ion{Ne}{9} and \ion{O}{7},
with evidence for similarly low $R$ ratios of \ion{N}{6} and \ion{Mg}{11}
\cite[]{kahn}. Direct observations of X-ray and UV spectra showed that
photoexcitation can account for the $R \lesssim 0.3$ ratio of \ion{N}{6} through
\ion{Ne}{9} in Hercules X-1, 
an intermediate-mass X-ray binary \cite[]{jimenez}. From our data
and that of C01, we have shown that in \exop, photoexcitation by the 
accretion disk UV emission can drive the $R \lesssim 0.2$ ratio of \ion{N}{6} through 
\ion{Mg}{11}.

\subsection{Elemental abundance ratios derived from edges}

While the Ne/O and Mg/O abundance ratios derived  
from our emission line model are within 3$\sigma$ of Solar, 
the edge-derived Mg/O ratio is larger than the Solar value.
The recombination line analysis implies that 
the EM decreases with $\xi$. If we assume the line-of-sight gas follows a similar trend,
then $N_{\rm Mg XI}/N_{\rm O VII} < {\rm Mg}/{\rm O}$ and
the abundance ratio ${\rm Mg}/{\rm O} > 4$
times the Solar value (3$\sigma$ limit). 
However, the edge-derived abundances may not be as reliable
as those obtained from emission lines, since the soft
continuum (absorbed by \ion{O}{7}) and the hard continuum (absorbed by \ion{Mg}{11}) 
require distinct spectral components, so our line of sight to them may differ.
There is a marginal detection of a larger-than-expected neutral
Mg K edge at 9.48~\AA \ in the spectrum of Figure \ref{fig:tot_detail}.
The abundance ratio obtained from the emission lines 
is ${\rm Mg}/{\rm O} = 10 \pm 5$ 
or ${\rm Mg}/{\rm O} = 16 \pm 9$ times Solar, depending
on the model chosen (Table \ref{tab:elem}).

\section{Conclusions}
\label{sec:concl}

The \chandra HETGS spectra reveal discrete absorption and emission
features which are the signatures of a photoionized plasma
in orbit around the neutron star in \exop.
We discern a compact source of continuum X-rays with hard and 
soft components, as well as 
an extended recombination emission region originating
in the outer decade in radius above the accretion disk.
The recombination emission is more prominent above 8~\AA \ and during dips,
since it is less absorbed than the compact continuum emission.
During three binary orbits,
the color dependent X-ray light curves show that
the intensity variations are due only to eclipses, type I bursts,
and intensity dips, with no evidence for accretion rate changes $> 10$\%.

The high-resolution X-ray spectra show
that intensity dips are caused by the increase in
column density of an absorber which is composed of photoionized plasma and neutral gas.
Both the \ion{Mg}{11} edge optical depth and \nH are correlated with dip activity.
The signatures of the photoionized absorber
are the K-shell edges of \ion{Mg}{11} and \ion{O}{7}, as well as
the \ion{O}{8} and \ion{O}{7} K-shell edges found by C01.
The periodic absorber is located in two bulges of the outer accretion
disk, one of which is near the accretion stream. Another component of
the absorber does not correlate with orbital period.
The photoionized medium also produces emission features
with a large velocity broadening of $500 < \sigma_v < 1200$\kmps, but with
a small systemic velocity $v < 310$\kmps (at 90\% confidence).
The radiative recombination emission lines \ion{Mg}{12} Ly$\alpha$,
\ion{Mg}{11} He$\alpha$, \ion{Ne}{10} Ly$\alpha$, \ion{Ne}{9} He$\alpha$, 
\ion{O}{8} Ly$\alpha$, and \ion{O}{7} He$\alpha$ are detected, as well as
the RRC of \ion{Ne}{10} and \ion{Ne}{9}, which are signatures of a
photoionized gas with $kT \lesssim 20$\eV. 
Our line model fits imply O/Ne and Mg/Ne abundance
ratios within 3$\sigma$ of the Solar values.
The ionized edges may, however, require a 
Mg/O abundance ratio $\gtrsim 4$ times Solar (3$\sigma$).

The photoionized plasma is located 
at the outer decade in radius of an accretion disk which
is thickened by X-ray illumination and by 
the impact of the accretion stream on the disk.
The plasma is located at a height $h \sim 0.2 r$ 
above and below the disk midplane, which
is larger than the \cite{ss73} model predictions.
Spectroscopic constraints on the ionization parameter,
density, kinematics, and UV radiation environment of the emitting plasma, 
all indicate that it is bound in Kepler orbits
inside the neutron star Roche Lobe.
X-ray photoionization heating 
from the neutron star and ram pressure from
the accretion stream can sufficiently 
expand the disk to explain the observed geometry.
Accretion disk atmosphere and corona models 
with vertically stratified ionization
can produce disks with the required height.
The ionization structure of the observed
emission region can be attributed to
vertical disk density gradients, with no
evidence for radial temperature stratification.
The measured spectra introduce stringent constraints on
the accretion disk structure, and further model comparisons are
warranted by the data.

\begin{figure}
\epsscale{.8}
\plotone{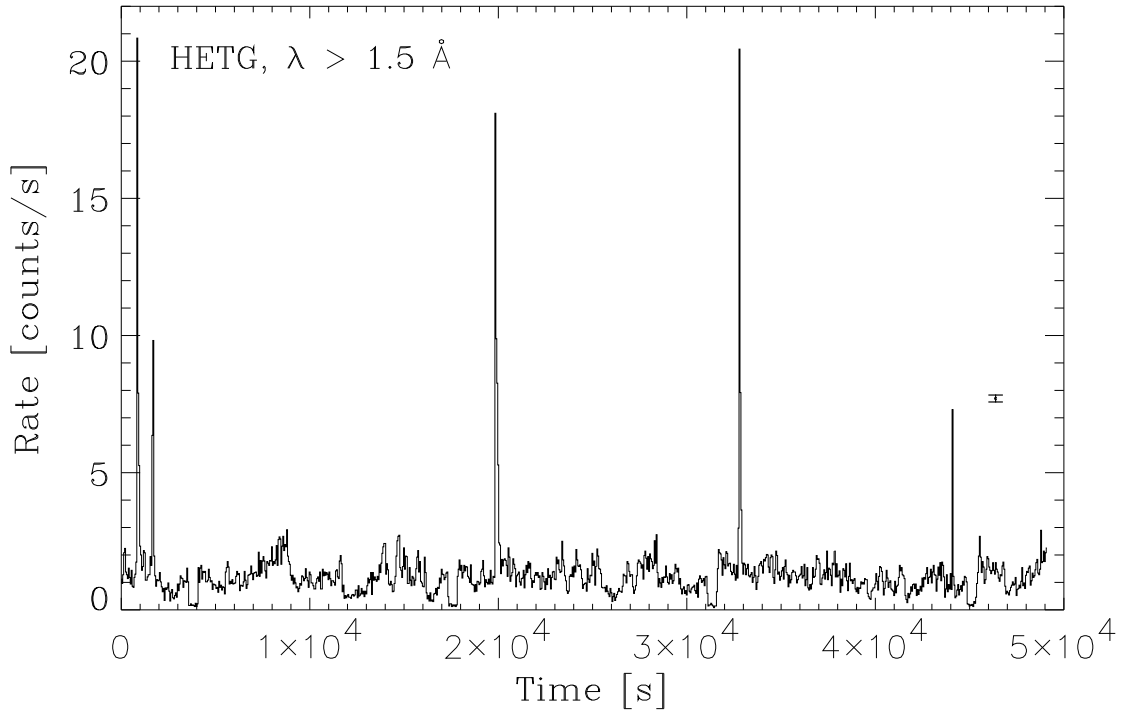}
\caption{Light curve of all dispersed photons in the HETGS from the
$1.5 < \lambda \lesssim 25$~\AA \ band in 50~s bins. The statistical
uncertainty at a rate of 1.0 count/s is shown. The light curve shows
sharp eclipse events, bright and short type I bursts, and intensity dips.
The decrease in brightness of type I bursts appears to be
associated with the dip events. 
\label{fig:totlc}}
\end{figure}

\begin{figure}
\epsscale{.8}
\plotone{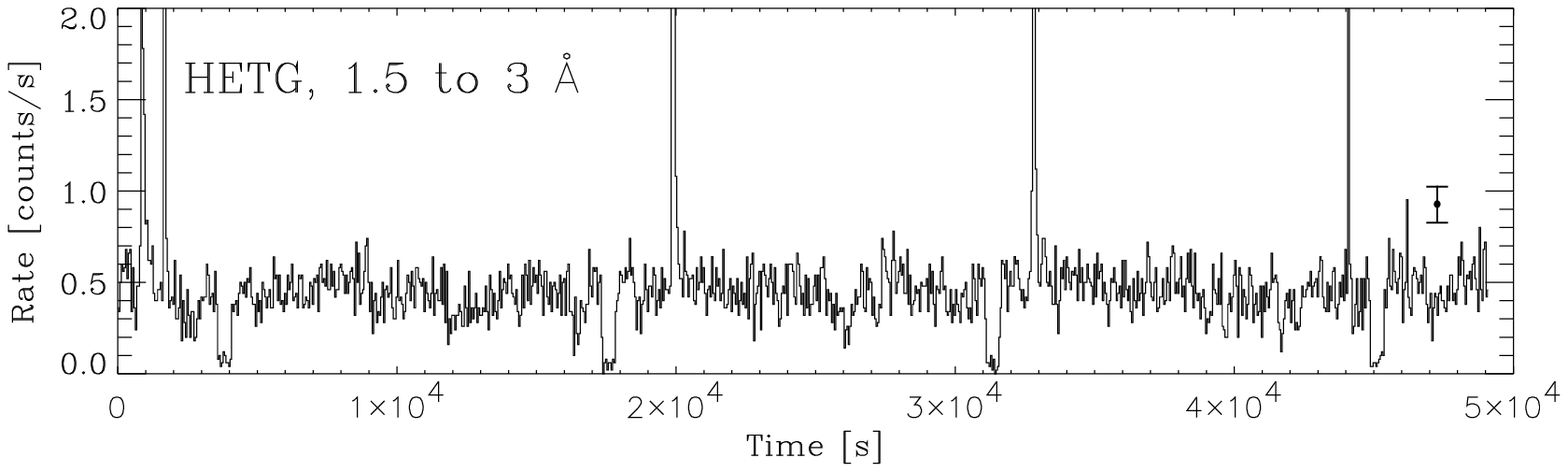}
\plotone{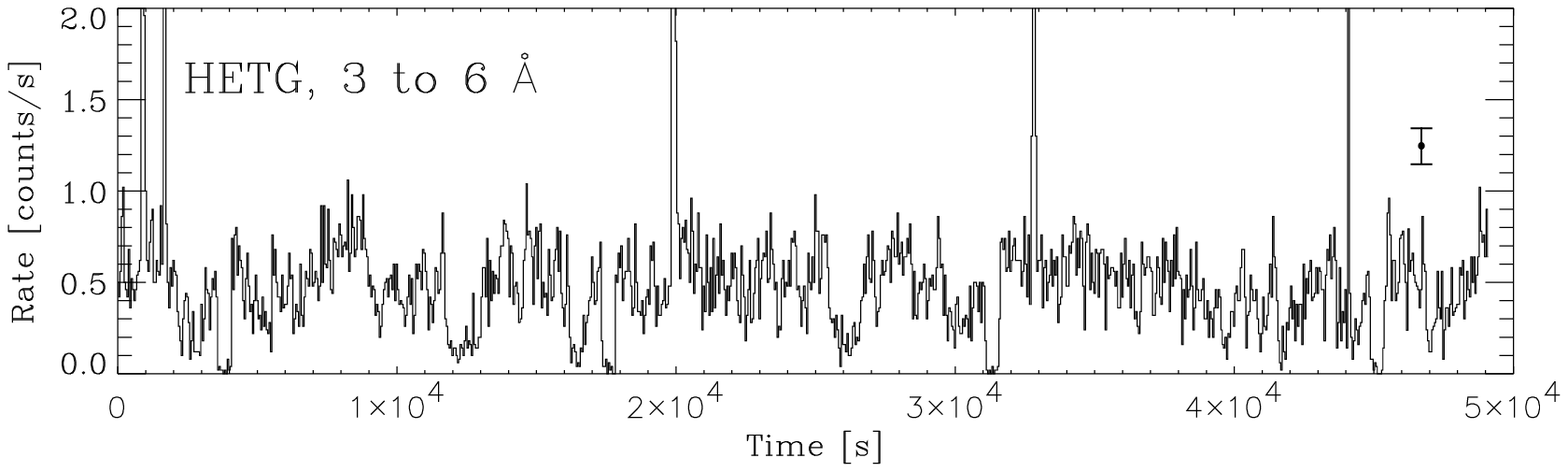}
\plotone{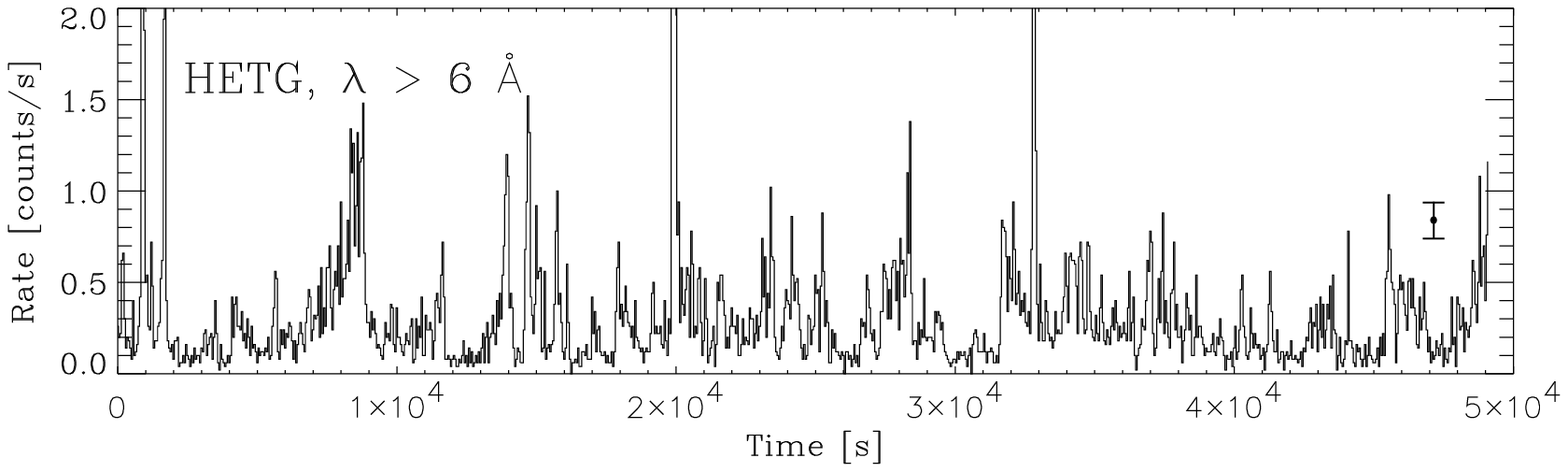}
\caption{Light curves for three distinct wavelength intervals in 50~s bins. 
The statistical uncertainty at a rate of 0.5 counts/s is shown. The
spectrum hardens during dip events, and the 1.3~\AA \ to 3~\AA \ light
curve implies a nearly constant accretion rate. The ``soft flare'' at 8~ks
shows the same color properties as the intervals without intensity dips.
\label{fig:wavelc}}
\end{figure}

We expect the spectra of the type I bursts in \exo to be partially
absorbed by the ionized plasma in the outer accretion disk, particularly
during the intensity dip phases and above $\sim 6$~\AA. 
The fifth burst in our HETGS observation shows evidence for
such absorption (see Figs. \ref{fig:totlc} and \ref{fig:wavelc}).
The association we found between dips and an ionized absorber 
implies that the ionized absorber column on the burst spectra 
will be largest at orbital phases $\phi \sim 0.6$ and $\phi \sim 0.9$, 
when a disk bulge is in the line of sight.
\cite{red} reported an ionized absorber outflowing during the late burst phases,
which they modeled in the spectrum to discern
the gravitationally redshifted features attributed to the neutron star
surface. Since the burst duration is much shorter than the outer disk dynamical timescale,
this ionized absorber outflow may be produced by
material in the outer accretion disk which is radiatively
driven by the burst. In that case, the nature of the outflow 
would be closely related to the structure of the accretion disk
and its time-dependent response to a type I burst.

\acknowledgments

We thank Prof. Claude Canizares, the P.I. of HETG, for providing these
data as part of his Guaranteed Target Observations (GTO).
We thank the members of the MIT CXC and HETG groups for their support.
Funding for this work was provided under NASA contract number NAS8-01129.

\begin{figure}
\epsscale{1}
\plotone{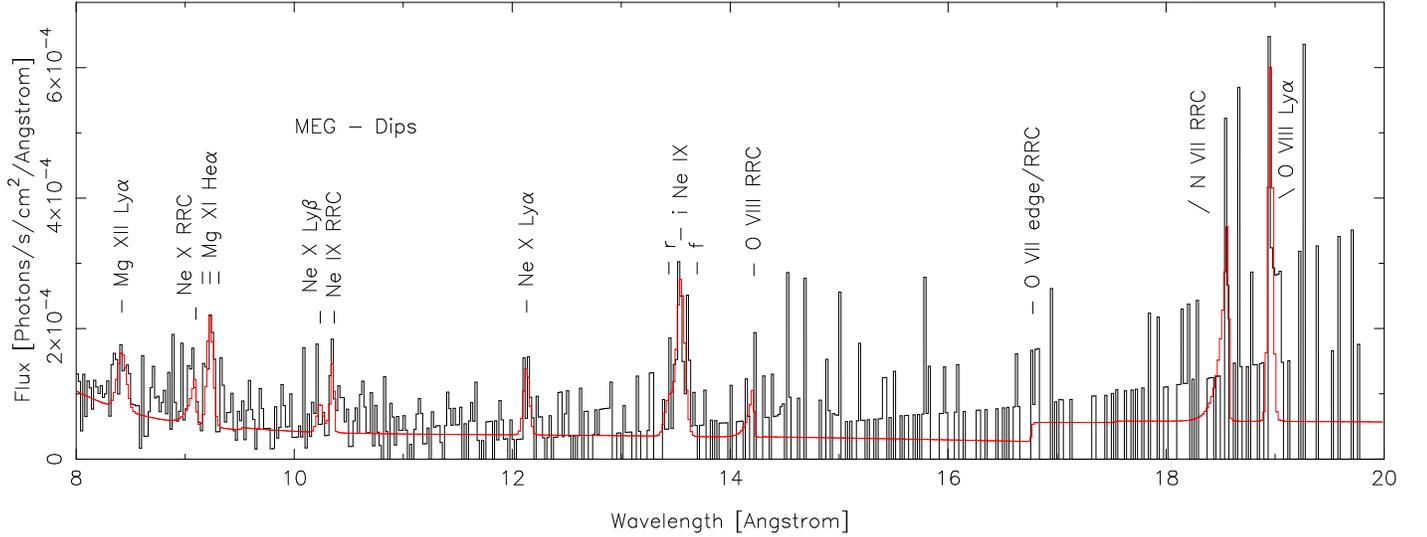}
\caption{Observed MEG spectrum during the dip state (black) and model fit (red).
The emission lines from H-like and He-like ions and their respective RRCs are
the signatures of a photoionized plasma. 
The \ion{Ne}{9} RRC and \ion{Ne}{10} RRC are used to constrain the
temperature (\S \ref{sub:temp}).
\label{fig:dip_detail}}
\end{figure}

\clearpage





\clearpage

\begin{deluxetable}{lrrrr}
\tabletypesize{\small}
\tablecaption{X-ray emission features
\label{tab:lines}}
\tablewidth{0pt}
\tablehead{
Feature & $\lambda$\tablenotemark{a}        & $\sigma_v$     & Flux    & State \\ 
 & [\AA]       & [\kmps]    & [$10^{-5}$ \phpcmsqps]   & \\ 
}
\startdata
O VII $f$       &     22.097    	& 700\tablenotemark{a} 	& $\le 2.4$         & Dip \\ 
	& 						& 700\tablenotemark{a}   & $\le 3.7$         & Persistent \\ 

O VII $i$      &     21.802    	& 700\tablenotemark{a}  & $4.7 \pm 1.9 $         & Dip \\ 
			&				& 700\tablenotemark{a}  & $5.9 \pm 3.4 $         & Persistent \\ 

O VII $r$       &     21.600    	& 700\tablenotemark{a} 	& $3.4 \pm 1.7 $         & Dip \\  
			&				& 700\tablenotemark{a} 	& $\le 1.4 $         & Persistent \\ 

O VIII Ly$\alpha$ &     18.970     & 700\tablenotemark{a} 	& $3.5 \pm 1.2 $         & Dip \\ 
			&				& 700\tablenotemark{a} 	& $3.7 \pm 2.0 $         & Persistent \\ 

N VII RRC	&      18.587 & 0\tablenotemark{a} & $     2.7 \pm    1.2 $         & Dip \\ 

Ne IX $f$       &     13.698    	& 1200\tablenotemark{a}       & $\le 0.4 $         & Dip \\ 
		&					& 500\tablenotemark{a}        & $\le 0.1 $         & Persistent \\ 

Ne IX $i$       &     13.552    	& $ 1200 \pm 330 $         & $2.46 \pm 0.61 $         & Dip \\ 
		&					& $ 500 \pm 270 $         & $2.21 \pm 0.83 $         & Persistent \\ 

Ne IX $r$       &     13.447    	& 1200\tablenotemark{a} & $\le 0.48 $         & Dip \\ 
			&				& 500\tablenotemark{a}  & $\le 2.40 $         & Persistent \\ 

Ne X Ly$\alpha$         &     12.135    & $570 \pm 250$         & $0.69 \pm 0.26 $         & Dip \\ 
					&			& 620\tablenotemark{a}  & $\le 1.04 $         & Persistent \\ 

Ne IX RRC&      10.370     & 0\tablenotemark{a} & $       0.49 \pm       0.18 $         & Dip \\ 

Ne X Ly$\beta$  &     10.239    	& 900\tablenotemark{a} & $0.35 \pm 0.16 $         & Dip \\ 
			&				& 732\tablenotemark{a} & $\le 0.82 $         & Persistent \\ 

Mg XI $f$       &      9.314    	& 900\tablenotemark{a} & $\le 0.36 $         & Dip \\ 
				&			& 650\tablenotemark{a} & $\le 0.44 $         & Persistent \\ 

Mg XI $i$       &      9.230     	& $890 \pm 220 $         & $1.18 \pm 0.20 $         & Dip \\ 
			&				& $650 \pm 330 $         & $1.05 \pm 0.41 $         & Persistent \\ 

Mg XI $r$       &      9.169     	& 900\tablenotemark{a} & $\le 0.43 $         & Dip \\ 
			&  				& 650\tablenotemark{a} & $\le 0.19 $  & Persistent \\ 

Ne X RRC&      9.102     & 0\tablenotemark{a} & $       0.52 \pm       0.17 $         & Dip \\ 

Mg XII Ly$\alpha$ 	&      8.419    	& 900\tablenotemark{a} 	& $0.62 \pm 0.22 $         & Dip \\ 
				&			& 750\tablenotemark{a} 	& $\le 0.31 $         & Persistent \\ 

Si XIV Ly$\alpha$       &      6.180    & 900\tablenotemark{a} & $ 0.51 \pm       0.26 $         & Dip \\ 
				&				& 750\tablenotemark{a} & $ \le 0.21 $         & Persistent \\

\enddata

\tablecomments{Errors and upper limits are 1$\sigma$.}
\tablenotetext{a}{These values are fixed.}
\end{deluxetable}

\clearpage

\begin{deluxetable}{cccccc}
\tablecaption{Diagnostics with He-like ion lines 
\label{tab:helike}} 
\tablewidth{0pt}

\tablehead{
		& Measured  	& Measured		& Derived & Derived &	\\
		& $R=f/i$  	& $G=(f+i)/r$ 		& $n_e^{\rm crit}$			& $F_\lambda^{\rm crit}$	&	\\
Ion 		& Line   		& Line  			&  [cm$^{-3}$]$^{(1)}$ 		& [\ergpcmsqpspA]		& State	\\
    		& Ratio  		& Ratio 			&  						& 	(est.)				&		\\ }
\startdata
Mg XI  	& $< 0.15$ 	& $> 3$ 	 	 	& $8 \times 10^{13}$	&  $2 \times 10^{9}$ 	& Dip \\ 	  
		& $< 0.4$ 	& $> 2$ 	 	 	& $3 \times 10^{13}$	& 	\nodata			& Persistent \\ 	  
Ne IX  	& $< 0.08$  	& $> 4$    		& $1 \times 10^{13}$	& \nodata		& Dip \\ 	  
		& $< 0.05$  	& $> 0.6$    		& $3 \times 10^{13}$	& $2 \times 10^8$ 	& Persistent \\ 	  
O VII  	& $< 0.3$  	& $2 \pm 1$  		& $3 \times 10^{11}$	& $6 \times 10^6$	& Dip \\ 	  
		& $< 0.4$  	& $> 2$ 	 		& $2 \times 10^{11}$	& \nodata			& Persistent \\ 	  
\enddata
\tablecomments{Errors and upper limits are 1$\sigma$. 
Symbols: $n_e^{\rm crit}$ = critical electron density, $F_\lambda^{\rm crit}$ = critical UV flux
at $f \to i$ transition wavelength.
C01 observed 
$R < 0.2$ for both \ion{O}{7} and \ion{Ne}{9}, which are
consistent with our HETGS observations.}
\tablerefs{(1) Porquet \& Dubau 2000.}
\end{deluxetable}

\begin{deluxetable}{lrrr}
\tablecaption{X-ray absorption features
\label{tab:edges}}
\tablewidth{0pt}
\tablehead{
Edge & $\lambda$ [\AA]        & 	$\tau$ 	& State \\ 
}
\startdata
Mg XI K & 7.037\tablenotemark{a} &      $0.36 \pm 0.07$ & Dip \\ 
&  &      $0.16 \pm 0.06$ & Persistent \\ 
O VII K & 16.78\tablenotemark{a} &   $0.76 \pm 0.13$  & Persistent \\ 

\enddata

\tablecomments{Errors are 1$\sigma$, with the continuum parameters and \nH \ set free.}
\tablenotetext{a}{These values are fixed.}
\end{deluxetable}

\begin{table}
\begin{center}
\small
\caption{Elemental abundance measurements and differential emission measure (DEM) 
parameters 
\label{tab:elem}}
\begin{tabular}{lccccc}
\\
\tableline\tableline
\underline{~$({\rm O}/{\rm Ne})$~} 	& \underline{~$({\rm Mg}/{\rm Ne})$~}  & $K$ 		& $-\gamma$ & $\chi^2$ of Fit & Ionizing \\
$({\rm O}/{\rm Ne})_{\sun}$ 		& $({\rm Mg}/{\rm Ne})_{\sun}$	& ($10^{58}$ cm$^{-3}$) & & (o.c.) & Spectrum \\
\tableline
$0.3 \pm 0.1$	 & $4.8 \pm 2.1$ 	& $4.4^{+1.6}_{-3.0}$  	& $1.2 \pm 0.4$ 	& 0.28 & 20~keV brems. \\
$0.4 \pm 0.15$ & $4.1 \pm 2.0$ 	& $0.85^{+0.30}_{-0.63}$	& $0.74 \pm 0.38$ 	& 0.52 & PL +cut \\
\tableline
\tableline
\end{tabular}
\tablecomments{Symbols: o.c. = over-constrained fit. We show the 1$\sigma$ statistical errors. The element 
abundance ratios (by number of atoms) are normalized to the solar values compiled by
\citet[]{wilms}, which are (O/Ne)$_{\sun} = 6.9$, and (Mg/Ne)$_{\sun} = 0.32$.
The $K$ and $\gamma$ parameters define the DEM. The abundance measurements depend
weakly on the assumed ionizing X-ray continuum.}
\end{center}
\end{table}


\begin{deluxetable}{lrrr}
\tabletypesize{\small}
\tablecaption{Continuum fits for two power law model 
\label{tab:contpl}}
\tablewidth{0pt}
\tablehead{ Fit parameter	& Units 			& Value  		& Value \\
						& 				& (Dips) 		& (Persistent) \\ }
\startdata
PL1 Norm. (1~keV) 	& phot~keV$^{-1}$ cm$^{-2}$ s$^{-1}$ & $(1.8 \pm 0.2) 10^{-2}$	& $(2.0 \pm 0.2) 10^{-2}$ \\
PL1 $N_{H}$  		& $10^{22}$ cm$^{-2}$  		& $4.2 \pm 0.1$ 		&  $2.4 \pm 0.2$  \\
PL1 Index 		& \nodata 				&  	$1.18 \pm 0.03$ 	&  $1.20 \pm 0.03$ \\
PL2 Norm. (1~keV) 	& phot~keV$^{-1}$ cm$^{-2}$ s$^{-1}$ & $(1.4 \pm 0.1) 10^{-3}$ 	&  $(2.4 \pm 0.2) 10^{-3}$ \\
PL2 $N_{H}$  		& $10^{22}$ cm$^{-2}$  		& $0.20 \pm 0.01$	&  $0.11$\tablenotemark{a}  \\
PL2 Index 		& \nodata 				&   3.3\tablenotemark{a}	&  $3.3 \pm 0.1$   \\
Cash-stat$/DOF$ 	& \nodata 				&   $2155/3072 = 0.70$ 	& $2317/3072 = 0.75$    \\

\enddata

\tablecomments{Errors and upper limits are 1$\sigma$.}
\tablenotetext{a}{These values are fixed.}
\end{deluxetable}

\begin{deluxetable}{lccc}
\tabletypesize{\small}
\tablecaption{Continuum fit parameters for partial covering model 
\label{tab:contpc}}
\tablewidth{0pt}
\tablehead{ Fit parameter	& Units 			& Value  		& Value \\
						& 				& (Dips) 		& (Persistent) \\ }
\startdata
BB Norm. (1~keV) & phot~keV$^{-1}$ cm$^{-2}$ s$^{-1}$ & $(9.3 \pm 0.9) 10^{-4}$ 	&  $9.3 \times 10^{-4}$~$^{a}$  \\
BB $kT$ 			& eV  					& 2$^{a}$			& 2$^{a}$	\\	
BB $N_{H}$  		& $10^{22}$ cm$^{-2}$  		& $23 \pm 2$		& $44 \pm 5$ \\
PL Cov. Frac. 		& \nodata 				& $0.976 \pm 0.001$ & $0.941 \pm 0.002$ \\
PL Index 			& \nodata 				& 1.7$^{a}$ 		& 1.7$^{a}$  \\
PL $N_{H}$  		& $10^{22}$ cm$^{-2}$  		& $4.60 \pm 0.03$	& $3.21 \pm 0.05$ \\
PL Norm. (1~keV) & phot~keV$^{-1}$ cm$^{-2}$ s$^{-1}$ & $(3.5 \pm 0.3) 10^{-2}$ & $(4.4 \pm 0.4) 10^{-2}$  \\
Cash-stat$/DOF$ 	& \nodata 				& $2073/3072 = 0.68$ 		& $2141/3072 = 0.70$  \\
\enddata

\tablecomments{Errors and upper limits are 1$\sigma$.}
\tablenotetext{a}{These values are fixed.}
\end{deluxetable}

\end{document}